\documentclass[twocolumn,twoside,slac_two]{revtex4}
\usepackage{graphicx}
\usepackage{fancyhdr}
\begin{document}

\title{Multiwavelength Spectral Studies Of Fermi-LAT Blazars}

%

\author{M. Joshi, A. Marscher, S. Jorstad}
\affiliation{Boston University, Boston, MA, USA}
\author{M. B\"{o}ttcher}
\affiliation{Ohio University, Athens, OH, USA}
\author{I. Agudo}
\affiliation{Boston University, Boston, MA, USA \& IAA, Granada, Spain}
\author{V. Larionov}
\affiliation{St. Petersburg State University, St. Petersburg, Russia}
\author{M. Aller}
\affiliation{University of Michigan, Ann Arbor, MI, USA}
\author{M. Gurwell}
\affiliation{SAO, Cambridge, MA, USA}
\author{A. L\"{a}hteenm\"{a}ki}
\affiliation{Mets\"{a}hovi Radio Observatory, Kylm\"{a}l\"{a}, Finland}

\begin{abstract}
We present multiwavelength spectral analyses of two Fermi-LAT blazars,
OJ 287 and 3C 279, that are part of the Boston University
multiwaveband polarization program. The data have been compiled from
observations with Fermi, Swift, RXTE, the VLBA, and various
ground-based optical and radio telescopes. We simulate the dynamic
spectral energy distributions (SEDs) within the framework of a
multi-slice, time-dependent leptonic jet model for blazars, with
radiation feedback, in the internal shock scenario. We use the
physical jet parameters obtained from the VLBA monitoring to guide our
modeling efforts. We discuss the role of intrinsic parameters and the
interplay between synchrotron and inverse Compton radiation processes
responsible for producing the resultant SEDs.

\end{abstract}

\maketitle

\thispagestyle{fancy}


\section{\label{intro}Introduction}
Blazar jets are highly violent in nature and are dominated by
ultrarelativistic particles. The SED of blazars consists of two
spectral bumps. The low-energy component is due to synchrotron
radiation emanating from relativistic particles, and the high-energy
component (for leptonic jet model) is a result of Compton upscattering
of the seed photon field by ultrarelativistic particles. The seed
photons could either be the synchrotron photons produced in the jet
(synchrotron self Compton, SSC) \citep[]{mg1985, gm1998}, and/or
external disk photons entering the jet directly (external Compton
disk, ECD) \citep[]{ds1993, bms1997}, and/or the disk photons getting
reprocessed in the broad line region (BLR) (external Compton cloud,
ECC) \citep[]{si1994, bb2000}, and/or the dusty torus \citep[]{bl2000,
  aps2002} and then entering the jet. The spectral variability patterns
and SEDs are important tools used for understanding the acceleration
of particles and the time-dependent interplay of various radiation
mechanisms responsible for the observed emission.

Here, we analyze the multiwaveband SED of two Fermi-LAT blazars, OJ287
and 3C 279, using the 1-D multi-slice time-dependent leptonic jet
model, with radiation feedback scheme of \cite{jb2011} to gain
understanding of the role of various intrinsic parameters and
radiation processes in producing the resultant SEDs.

We briefly describe the model of \cite{jb2011} in \S \ref{model}. We
discuss our first results from this study in \S \ref{results}. We
summarize our results and give a brief description of future work in
\S \ref{summary}.

\section{\label{model}Internal Shock Model}
The mode of acceleration of plasma electrons (and positrons) to highly
relativistic energies and its location in the jet is still not
completely understood. One way to comprehend the physics of particle
acceleration is the internal shock model, in which the central engine
(black hole + accretion disk) spews out shells of plasma with
different velocity, mass, and energy. The collision between such
shells gives rise to internal shocks (reverse (RS) and forward (FS)),
which convert the ordered bulk kinetic energy of the plasma into the
magnetic field energy and random kinetic energy of the particles. The
highly accelerated particles then radiate and produce the emission
observed from the jet.

\begin{figure}
\includegraphics[width=65mm]{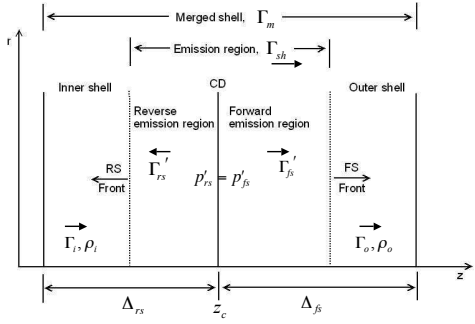}
\caption{Schematic of the emission region with RS traveling into the
  inner shell of bulk Lorentz factor (BLF) $\Gamma_{\rm i}$, and FS
  moving into the outer shell of BLF with $\Gamma_{\rm o}$, such that
  $\Gamma_{\rm i} > \Gamma_{\rm o}$. The primed quantities refer to
  the comoving frame and unprimed refer to the lab (AGN) frame. The
  comoving pressures $p_{\rm fs}^\prime$ and $p_{\rm rs}^\prime$ of
  the shocked fluids across the contact discontinuity (CD) are equal.
  $\Delta_{\rm rs}$ and $\Delta_{\rm fs}$ are the widths of the inner
  and outer shell after the collision obtained from the shock dynamics
  \citep{sp2001}.}
\label{emission_region}
\end{figure}

The collision of two plasma shells results in an emission region as
shown in Figure \ref{emission_region}. The treatment of shell
collision and shock propagation is hydrodynamic and relativistic in
nature \citep{sp2001}.

The evolution of electron and photon population inside the emission
region are governed, respectively, by,

\begin{equation}
\label{2}
{\partial n_{e} (\gamma, t) \over \partial t} = -{\partial \over
  \partial \gamma} \left[\left({d\gamma \over dt}\right)_{loss} n_{e}
  (\gamma, t)\right] + Q_{e} (\gamma, t) - \frac{n_{e} (\gamma,
  t)}{t_{e,esc}}
\end{equation}

and

\begin{equation}
\label{3}
{\partial n_{ph} (\epsilon, t) \over \partial t} = \dot n_{ph,em}
(\epsilon, t) - \dot n_{ph,abs} (\epsilon, t) - \frac{n_{ph}
  (\epsilon, t)}{t_{ph,esc}}
\end{equation}

Here, $(d\gamma/dt)_{\rm loss}$ is the radiative energy loss rate, due
to synchrotron and SSC losses, for the electrons. $Q_{e} (\gamma, t)$
is the sum of external injection and intrinsic $\gamma - \gamma$ pair
production rate and $t_{\rm e,esc}$ is the electron escape time scale.
$\dot n_{\rm ph,em} (\epsilon, t)$ and $\dot n_{\rm ph,abs} (\epsilon,
t)$ are the photon emission and absorption rates corresponding to the
electrons' radiative losses, and $t_{\rm ph,esc} = (3/4)R_{b}/c$ is
the photon escape timescale. The evolution of the electron and photon
population is followed in a time-dependent manner inside the emission
region and radiative energy loss rates as well as photon emissivities
are calculated using the time-dependent radiation transfer code of
\cite{jb2011}.

The model follows the evolution of the emission region out to sub-pc
scales and simulates only the early phase of $\gamma$-ray
production. During this time, the radiative cooling is strongly
dominant over adiabatic cooling and the emission region is highly
optically thick out to GHz radio frequencies. Thus, the simulated
radio flux is well below that of the actual radio data. Also, the
phase of the emission region in which it gradually becomes transparent
to radio frequencies is not simulated, as that would require the
introduction of several additional, poorly constrained parameters.

\subsection{\label{scheme}Multi-slice Radiation Transfer Scheme}
\begin{figure}
\includegraphics[width=65mm]{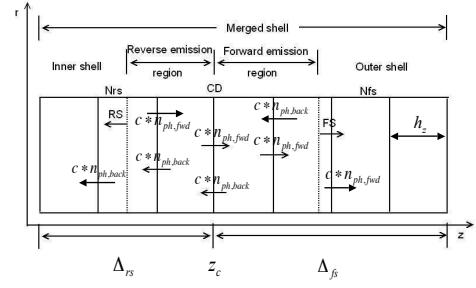}
\caption{Schematic of the radiative transfer in between the slices
  using the appropriate photon escape probability function. The
  unprimed $n_{\rm ph, \sim fwd/back}$ values represent the photon
  densities in the forward and backward direction, respectively, in
  the comoving frame of the emission region. The rest of the unprimed
  quantities refer to the lab frame.}
\label{zone_dia}
\end{figure}

A cylindrical emission region is considered to calculate the resultant
spectrum in a time-dependent manner. The inhomogeneity in the photon
and particle density throughout the emission region is realized by
dividing the region into multiple slices, as shown in Figure
\ref{zone_dia}.

The photon density of a zone and the probability of escape for a
photon from that zone in a particular direction (forward, backward, or
sideways), P, are used to calculate photon escape rates in that
direction, according to
\begin{equation}
\label{ndoteqn}
\frac{dn_{\rm ph, fwd/back/side}(\epsilon, \Omega)}{dt} = \frac{n_{\rm                    
    ph}(\epsilon, \Omega)}{t_{\rm ph, esc}} P_{\rm fwd/back/side}
\end{equation}
where $t_{\rm ph, esc}$ is the photon escape timescale for a
cylindrical region.

We use the scheme presented in equation \ref{ndoteqn} to calculate the
radiation transfer within each slice and in between the slices
\citep{jb2011}.

\section{\label{results}First Results}
The time-dependent model, calculating synchrotron and SSC radiation
processes, developed in \cite{jb2011} has been used to reproduce the
observed SED of two Fermi-LAT blazars, OJ287 and 3C~279 that are part
of the Boston University multiwaveband monitoring program. We have
collected the data from observations with Fermi, Swift, RXTE, the
VLBA, and various ground-based optical and radio telescopes to
construct the multiwaveband SEDs of OJ287, and 3C~279, the latter of
which corresponds to the quasar's optical high state as observed on 15
January 2006.

\begin{figure}[ht]
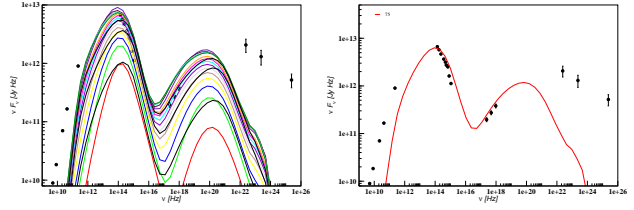

\begin{center}$
\begin{array}{cc}
\includegraphics[width=40mm]{f3.eps} &
\includegraphics[width=40mm]{f4.eps}
\end{array}$
\end{center}
\caption{Simulated instantaneous and time-integrated (averaged over
  1 day) SED of OJ287 for 10/28/2008.}
\label{oj287}
\end{figure}

OJ287 is a BL Lac object purported to have a black hole binary system
\citep{vlp1999}. The blazar has exhibited spectral and polarization
variability in the past \citep{ai2011}. Here, we apply the model of
\cite{jb2011} with the ECD component included (Joshi et al., 2012, in
prep.) with 50 slices in the forward and 50 slices in the reverse
emission regions to analyse the SED of OJ287 as observed on
10/28/2008. Figure \ref{oj287} shows the instantaneous and
time-integrated simulated SED of OJ287 for that day. The model
independent parameters \citep{bh2005} estimated from the SED, VLBA
observations \citep{jo2005}, and variability on 1-day timescale were
used to develop an initial set of input parameters:

\begin{eqnarray}
\delta &\approx& 16.5 \hss \cr
R &\approx& 3.5 \times 10^{15} \; {\rm cm} \hss \cr
B &\approx& 0.52 \, \epsilon_B^{2/7} \; {\rm G} \hss \cr
\gamma_{\rm min} &\approx& 1.1 \times 10^3 \hss \cr
\gamma_{\rm max} &\approx& 3.0 \times 10^4 \hss \cr
q &\approx& 4.2 \hss \cr
\theta_{\rm obs} &\approx& 3.2^{o} \hss
\label{param_sumoj}
\end{eqnarray}

Here, $\delta$ is the Doppler boosting factor assumed to be equal to
the BLF of the emission region (as estimated from VLBA
observations). The symbols R \& B represent the values of comoving
radius and magnetic field in the emission region and $\epsilon_{\rm
  B}$, which is not a part of the model of \cite{jb2011}, is the ratio
of magnetic field and electron energy density assumed to be equal to 1
here. The $\gamma_{\rm min}$ \& $\gamma_{\rm max}$ refer to the low
and high energy cutoffs of the electron energy distribution. The
spectral index of the electron population is given by q and
$\theta_{\rm obs}$ is the observing angle inferred from VLBA
observations.

The initial set of parameters was modified to reproduce the state of
OJ287 as observed on 10/28/2008. Table~\ref{paramlist} lists the
parameters used for obtaining the resultant SED of the source shown in
Fig. \ref{oj287}. These parameters result in a $\Gamma_{\rm sh}
\approx 16.4$ for the entire emission region, $B \approx 5.0 ~G$ and
$\gamma_{\rm max} \approx 1.5 \times 10^5$ for both forward and
reverse emission regions, and $\gamma_{\rm min, ~fs} \approx 5.6
\times 10^2$ \& $\gamma_{\rm min, ~rs} \approx 1.0 \times 10^3$ for
forward and reverse emission regions, respectively.

As can be seen from the figure, the lower-energy bump of the
time-integrated simulated SED passes very close to the IR, optical,
and UV data points, indicating that the synchrotron component is
responsible for this part of the jet emission. The spectral upturn
takes place in the soft X-rays at $\geq$ 0.14 keV due to the presence
of the SSC component in the simulation and the lower-energy part of
the SSC component passes close to the X-ray data implying the
dominance of SSC component in producing this part of the high-energy
bump. The model (SSC+ECD), at this point, underpredicts the
$\gamma$-ray photon flux suggesting that the contribution from the BLR
might play a dominant role in reproducing this emission.

\begin{figure}[ht]
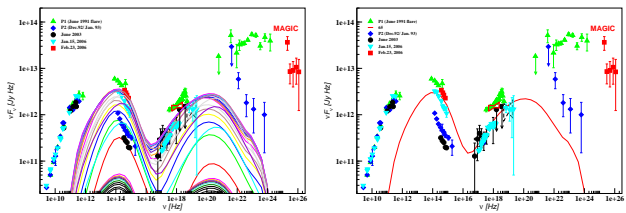

\begin{center}$                                         
\begin{array}{cc}  
\includegraphics[width=40mm]{f5.eps} &
\includegraphics[width=40mm]{f6.eps}
\end{array}$
\end{center}
\caption{Simulated instantaneous and time-integrated (averaged over
  1 day) SED of 3C~279 for 01/15/2006.}
\label{3c279}
\end{figure}

The flat-spectrum radio quasar (FSRQ), 3C~279, was observed in its
optical high state on 01/15/2006, almost a month before it was
observed by MAGIC emitting in the TeV energy regime for the first time
ever \citep{co2010}. Figure \ref{3c279} shows the instantaneous and
time-integrated simulated SED of 3C~279 for that day. The model
independent parameters \citep{bh2005} that were estimated using the
SED, VLBA observations \citep{jo2005}, and variability on 1-day
timescale were used to develop an initial set of input parameters:

\begin{eqnarray}
\delta &\approx& 15.5 \hss \cr
R &\approx& 2.5 \times 10^{16} \; {\rm cm} \hss \cr
B &\approx& 0.82 \, \epsilon_B^{2/7} \; {\rm G} \hss \cr
\gamma_{\rm min} &\approx& 9.0 \times 10^2 \hss \cr
\gamma_{\rm max} &\approx& 2.2 \times 10^4 \hss \cr
q &\approx& 4.3 \hss \cr
\theta_{\rm obs} &\approx& 2.1^{o} \hss
\label{param_sum3c}
\end{eqnarray}

All symbols refer to the same quantites as explained above and the
entire emission region, as mentioned above, has been divided into 100
slices to analyze the observed SED of the source. The initial set of
parameters was modified to reproduce the state of 3C~279 as observed
on 01/15/2006. Table~\ref{paramlist} lists the parameters used for
obtaining the resultant SED of the source shown in
Fig. \ref{3c279}. These parameters result in $\Gamma_{\rm sh} \approx
16.6$ for the entire emission region, $B \approx 4.0 ~G$ and
$\gamma_{\rm max} \approx 2.2 \times 10^5$ for both forward and
reverse emission regions, and $\gamma_{\rm min, ~fs} \approx 5.8
\times 10^2$ \& $\gamma_{\rm min, ~rs} \approx 1.3 \times 10^3$ for
forward and reverse emission regions, respectively.

As can be seen from the figure, the observed SED for 01/15/2006 shows
a high-energy bump that is indicative of a dominant SSC component and
a suppressed EC component. The time-integrated simulated SED passes
very close to the IR and optical data points, indicating that the
synchrotron component responsible for the lower energy bump of the
SED. The spectral upturn takes place in the soft X-rays at $\geq$ 0.11
keV due to the presence of the SSC component in the simulation. The
lower-energy part of the SSC component reproduces the X-ray data quite
well, suggesting the dominance of SSC component in producing this part
of the high-energy bump.
\\

\begin{table}[ht]
\begin{center}
\caption{Model parameters used to reproduce the state of OJ287 and 3C~279
  as observed on 10/28/2008 \& 01/15/2006, respectively.\\}
\begin{tabular}{|l|c|c|c|c|c|c|c|}
\hline \textbf{Source} & \textbf{$L_{w}$} & \textbf{$\Gamma_{i}$} & \textbf{$\Gamma_{o}$} & \textbf{q} & \textbf{$\varepsilon_{e}$} & \textbf{$\varepsilon_{B}$} & \textbf{$\zeta_{e}$}
\\
\hline & [$10^{47}$~ergs/s] & & &  & [$10^{-2}$] & [$10^{-3}$] & [$10^{-2}$]\\
\hline 3C279  &  5  &  35   &  10   &  4    &  9   &  3   &  2.5\\
\hline OJ287  &  4  &  25   &  12   &  4.2  &  9   &  14   & 1.0\\  
\hline
\end{tabular}
\\
\begin{tabular}{|l|c|c|c|c|c|}
\hline \textbf{Source} & \textbf{R} & \textbf{$\theta_{\rm obs}$} & \textbf{$L_{\rm disk}$} & \textbf{$M_{\rm BH}$} & \textbf{$\eta_{\rm acc}$}
\\
\hline & [$10^{16}$~cm] & [deg] & [$10^{44}$~ergs/s] & [$10^{8} M_{\rm Sol}$] & [$10^{-2}$]\\
\hline 3C279  &  3.7  &  2.5   &  -  &   -  &  -\\
\hline OJ287  &  4.0  &  2.5   &  2  &   2  &  6\\
\hline
\end{tabular}
\\

  $L_{w}$: luminosity of the injected electron population in the blob,
  $\Gamma_{i,o}$: BLFs of the inner and outer shells before collision,
  q: particle spectral index, $\varepsilon_{e}$: ratio of electron and
  shock energy density, $\varepsilon_{B}$: ratio of magnetic field and
  shock energy density, $\zeta_{e}$: fraction of accelerated
  electrons, R: comoving radius, $\theta_{obs}$: viewing angle,
  $L_{\rm disk}$: accretion disk luminosity, $M_{\rm BH}$: Mass of the
  BH, and $\eta_{\rm acc}$: accretion efficiency
\label{paramlist}
\end{center}
\end{table}

\section{\label{summary}Discussion and Future Work}
The time-integrated SEDs of both OJ287 \& 3C~279 need further
adjustments of parameters that are listed in Table~\ref{paramlist} in
order to obtain a satisfactory fit. The external Compton component due
to photons entering the jet from the BLR and dusty torus needs to be
incorporated in the existing model of \cite{jb2011} to correctly
reproduce the SEDs of blazars, especially for flat spectrum radio
quasars (Joshi et al. 2012, in prep.).

Further, we plan to incorporate the effects of magnetic field
orientation, as inferred from polarization monitoring programs, on the
resultant spectral variability and SEDs of blazars. This would further
aid us in the study of intrinsic parameter differences between various
blazar subclasses, arising from the orientation of the magnetic field
in the jet.

We plan to study the evolution of SED of blazars from quiescent to
flaring state in the light of the modified time-dependent model of
\cite{jb2011} (Joshi et al. 2012, in prep.). We will then compare the
results with multi-waveband data gathered by the Fermi-LAT and other
telescopes.

\bigskip 
\begin{acknowledgments}
This research was supported in part by NASA through Fermi grants
NNX10AO59G, NNX08AV65G, and NNX08AV61G and ADP grant NNX08AJ64G, and
by NSF grant AST-0907893.
\end{acknowledgments}

\bigskip 

\end{document}